\magnification=1200
\baselineskip=16pt
\nopagenumbers
\headline={\hss\tenrm -- \sec\folio --\hss}

\def\halv{{1\over 2}}
\def\p2{2\pi}

\def\gp2{{1\over \p2}}
\def\partx{{\partial \over\partial x}}
\def\ipartx{\partial /\partial x}
\def\party{{\partial \over\partial y}}
\def\partz{{\partial \over\partial z}}
\def\partt{{\partial \over\partial t}}

\def\con{\;{\rm const.}\;}

\def\snok2#1{\hbox{$\tilde{\mkern -2.0mu \tilde #1} $}}

\def\sec{}
\line{}
\bigskip
\centerline{\bf Well-posed initial-boundary value problem for the
harmonic Einstein equations}
\centerline{\bf using energy estimates}

\medskip\centerline{by}
\centerline{H.-O. Kreiss${}^{1,2}$, O. Reula ${}^{3}$, O. Sarbach${}^{4}$
and J. Winicour${}^{2,5}$ }
\medskip

${}^{1}$NADA, Royal Institute of Technology, 10044 Stockholm, Sweden

${}^{2}$Albert Einstein Institute, Max Planck Gesellschaft,
Am M\"uhlenberg 1, D-14476 Golm, Germany

${}^{3}$FaMAF Universidad Nacional de Cordoba,
Cordoba, Argentina 5000

${}^{4}$Instituto de F{\'{\i}}sica y Matem{\'{a}}ticas,
Universidad Michoacana de San Nicol{\'{a}}s de Hidalgo,
Edificio C-3, C.~P. 58040 Morelia, Michoac{\'{a}}n, M{\'{e}}xico  

${}^{5}$Department of Physics and Astronomy,
University of Pittsburgh, Pittsburgh, Pennsylvania 15260, USA

\bigskip
\bigskip

\centerline{\bf Abstract}

\bigskip
\noindent
In recent work, we used pseudo-differential theory to establish
conditions that the initial-boundary value problem for second order
systems of wave equations be strongly well-posed in a generalized sense.
The applications included the harmonic version of the Einstein equations.
Here we show that these results can also be obtained via standard energy
estimates, thus establishing strong well-posedness of the harmonic
Einstein problem in the classical sense.


\bigskip
\medskip\noindent
{\bf 1. Introduction}

\bigskip

In the harmonic description of general relativity, the Einstein equations
reduce to a constrained system of 10 quasilinear wave equations for the
components of the spacetime metric. Recently [1] we used the theory of
pseudo-differential operators to prove that one can construct constraint
preserving boundary conditions of Sommerfeld type such that the resulting
initial-boundary value problem (IBVP) is well-posed in the generalized
sense. We show in this paper that the
decisive estimate can also be obtained by {\it integration by parts}. It
follows that the full quasilinear system can be treated by standard
energy estimates to establish that the harmonic IBVP is strongly
well-posed in the classical sense [2].

Our results have broad application to other systems of second order wave
equations besides general relativity, e.g. to elasticity theory, acoustics and
electromagnetic theory. Most analytic and computational treatments of the IBVP
utilize the well-developed theory of first order symmetric hyperbolic systems.
We develop our results here in their natural second order form, which
avoids the integrability constraints associated with the extra
variables introduced in a reduction to first order.

In view of the wide range of potential applications, instead of the geometrical
notation of general relativity, we present our main results in Sec's 2 - 3 in a
style familiar to a broad audience of computational mathematicians and
physicists.  We use the notation and definitions of the classic treatise [2] on
the IBVP. In these sections, we treat systems of wave equations with constant
coefficients, as arise in the frozen coefficient form of the harmonic Einstein
equations. In Appendix 1, we
show that our results extend locally in time to the well-posedness of
quasilinear problems, such as the harmonic IBVP. 

There is an intimate interplay between the treatment of the analytic theory
using energy estimates and its finite difference approximation [2,3]. The proof
of existence of analytic solutions outlined in Appendix 1 is based upon a
convergent finite difference approximation incorporating semi-discrete
energy estimates via {\it summation by parts}. Conversely, using summation
by parts, the energy estimates of the analytic theory can be parroted by
discrete energy estimates which guarantee the stability of a finite difference
approximation [3]. In previous treatments of the second order wave equation,
the discrete energy approach has been used to develop stable difference
algorithms for Neumann and Dirichlet boundary conditions [4,5] and this
treatment has been extended to the Einstein equations [6]. The results
presented here provide a guide for applying the discrete energy approach to
second order systems with a wide range of boundary conditions. These include
the Sommerfeld condition, which has important application to outgoing
wave problems, but also more complicated conditions involving
derivatives tangential to the boundary.

The set of constraint-preserving boundary conditions allowed for the Einstein
system is quite extensive. The selection of a ``preferred'' choice, e.g.
Dirichlet, Neumann or Sommerfeld, rests upon additional geometrical or  physical
criteria depending upon the nature of the problem. An important example is the
description of an isolated radiating system, for which the physically
appropriate boundary condition should be adapted to the absence of external
influences. Unless the treatment of such a system is extended to infinity, this
requires introduction of an artificial boundary for which boundary data may not
be known, so that it becomes advantageous to use a boundary condition for which
homogeneous data is a good approximation. We defer this issue to future work
where we will present a geometric version of the results of this paper to
formulate boundary conditions appropriate for an isolated gravitational system.
Additional subtleties arise in the treatment of boundaries which are moving,
such as an oscillating conducting boundary in an electromagnetic wave problem or
the artificial boundaries that arise in the dynamically curved spacetime of
general relativity. In Appendix 2, we illustrate for the wave equation on a
curved space background how the geometric approach can be used to simplify the
formulation of energy estimates for such moving boundaries.

\bigskip\noindent
{\bf 2. The main estimate}
\bigskip

Consider the wave equation
$$ u_{tt}=u_{xx}+u_{yy}+u_{zz}+F \eqno(1) $$
on the half-space
$$ x\ge 0,\quad -\infty <y<\infty,\quad -\infty < z < \infty, $$
with boundary conditions
$$ \alpha u_t=u_x+\beta_1 u_y+\beta_2 u_z+\alpha q,\quad {\rm for}\quad 
x=0,~\alpha >0,~\beta_1^2+\beta_2^2<\alpha^2 , \eqno(2)
$$
with boundary data $q$ and initial data
$$ u=f_1,\quad u_t=f_2,\quad t=0 \eqno(3) $$
of compact support.
The subscripts $(t,x,y,z)$ denote partial derivatives,
e.g $u_t={\partial u\over \partial t}$. We assume that all
coefficients, the data and the solution are real and
that $\alpha >0,~\beta_j$ are constants. Also, we use the notation
$$ 
(u,v), \quad \|u\|^2=(u,u);\quad (u,v)_B,\quad
\|u\|^2_B=(u,u)_B,
$$
to denote the $L_2$-scalar product and norm over the half-space and 
boundary space, respectively.

Strong well-posedness of an IBVP extends the requirements of a well-posed
Cauchy problem to include estimates of boundary values [2]. In order to
adapt the standard definition to second order systems, we write
${\bf u}= (u,u_t,u_x,u_y,u_z)$ to represent the solution $u$ and
its derivatives; and similarly for the initial data we write
${\bf f}_1= (f_1,f_{1x},f_{1y},f_{1z})$.
For the problem (1)-(3), strong well-posedness requires the existence of a
solution satisfying the estimate
$$\|{\bf u(t)}\|^2+\int_0^t\|{\bf u(s)}\|^2_B ds 
       \le K_T \left (\|{\bf f}_1\|^2 +\|f_2\|^2+\int_0^t\|F(s)\|^2 ds
       +\int_0^t\|q(s)\|^2_B ds \right ),$$
where for every finite  time interval $0\le t \le T$ the constant $K_T$ is
independent of $F$, $f_1$, $f_2$ and $q$.

In [1] we used pseudo-differential theory to prove that the problem (1)-(3)
is well-posed in the generalized sense and that it is boundary stable. 
The proof in [1] is only given in 2D. In 3D the necessary and sufficient
condition for $\alpha$ is  $\alpha >\sqrt{\beta_1^2+\beta_2^2}.$ This same
inequality governs the strong well-posedness of the problem. (It corresponds to
the timelike property of the vector field $T^b$ in the geometric treatment of
Appendix 2.)

We now want to prove that results can also be obtained in terms of standard
estimates using integration by parts. For simplicity of presentation, we
restrict our attention here to estimates of the derivatives of $u$. An
estimate for $u$ itself can easily be obtained by the change of variable
$u\to e^{\gamma t}u$, as described in Appendix 1. We start with
\proclaim
Lemma 1. Let $\gamma_1,\gamma_2$ be real and $\gamma_1^2+\gamma_2^2<1.$
Then
$$
E=\|u_t\|^2+\|u_x\|^2+\|u_y\|^2+\|u_z\|^2-
2(u_t,\gamma_1 u_y+\gamma_2 u_z) \eqno(4)
$$
is a norm for the derivatives $(u_t,u_x,u_y,u_z)$.
\par
\medskip\noindent
{\it Proof.} Since $\gamma_1^2+\gamma_2^2< 1,$
we can choose $\delta$ with $0<\delta< 1$ such that
$$\eqalign{
2|(u_t,\gamma_1 u_y+\gamma_2 u_z)|&\le (1-\delta)\|u_t\|^2+
{1\over 1-\delta}(\gamma_1^2+\gamma_2^2)
(\|u_y\|^2 +\|u_z\|^2)\cr
&\le
 (1-\delta)\left(\|u_t\|^2+ \|u_y\|^2+\|u_z\|^2\right).\cr}
$$
This proves the lemma.
\medskip

Let $\gamma_1={\beta_1\over \alpha},~ \gamma_2={\beta_2\over \alpha}$.
Now we can prove that there is a standard energy estimate.
\proclaim
Theorem 1. 
The solution of  (1)--(3) satisfies the energy estimate
$$
\partt E  +{1\over \alpha} \|u_x\|^2_B\le E + \|F\|^2 +\alpha \|q\|^2_B. \eqno(5)
$$

\par
\medskip\noindent
{\it Proof.} Integration by parts gives us
$$
\partt \|u_t\|^2=2(u_t,u_{tt})=-\partt (\|u_x\|^2+\|u_y\|^2+\|u_z\|^2)+
2(u_t,F)-2(u_t,u_x)_B  \eqno(6)
$$
$$ \eqalign{
2\partt (u_t,\gamma_1 u_y+\gamma_2 u_z)&=
2(u_{tt},
\gamma_1u_y+\gamma_2u_z)\cr
&= -2(u_{x},\gamma_1 u_y+\gamma_2 u_z)_B 
      +2 (F, \gamma_1 u_y+\gamma_2 u_z),\cr}
\eqno(7)$$
where, for example, we have used $(u_{yy},u_z)=-(u_y,u_{yz})=0$.
Since (2) implies
$$
2(u_t,u_x)_B={2\over\alpha}\|u_x\|^2_B+
2(u_{x},\gamma_1 u_y+\gamma_2 u_z)_B+
2(u_x,q)_B,$$
by subtracting (7) from (6) we obtain
$$ \eqalign{
   \partt E &= 2(u_t-\gamma_1 u_y -\gamma_2 u_z,F)-{2\over\alpha}\|u_x\|^2_B
      -2(u_x,q)_B  \cr
         & \le \|u_t-\gamma_1 u_y -\gamma_2 u_z\|^2 + \|F\|^2 
       -{1\over \alpha} \|u_x\|^2_B +\alpha \|q\|^2_B. \cr}
$$
The identity
$$
  \| u_t -\gamma_1 u_y -\gamma_2 u_z\|^2 = E-\| u_x\|^2 -\| u_y\|^2 
   -\| u_z\|^2  + \| \gamma_1 u_y+\gamma_2 u_z \|^2
$$
then implies (5) and thus proves the theorem.

The theorem tells us that we can estimate 
$$
E(T)~\hbox{and}~\int_0^T \|u_x\|^2_B dt \quad \hbox{in terms of}\quad
E(0),~\int_0^T \|F\|^2dt~ \hbox{and}~
\int_0^T \|q\|^2_B dt.
$$
For the application to the Einstein equations we also need estimates
of the boundary norms of $u_y$ and $u_z$. We have
$$ \eqalign{
\partt (u_x,u_t)&=(u_{xt},u_t)+(u_x,u_{tt})\cr
&=-\halv\|u_t\|^2_B+(u_x,u_{xx})+(u_x,u_{yy})+(u_x,u_{zz})+(u_x,F)\cr
&=-\halv \|u_t\|^2_B-\halv\|u_x\|^2_B+\halv\|u_y\|^2_B+\halv\|u_z\|^2_B
+(u_x,F) .\cr}
\eqno(8) $$
The boundary conditions (2) give us, for any $\delta$ with $0<\delta<1,$
$$ \eqalign{
\|u_t\|^2_B&= \|\gamma_1 u_y+ \gamma_2 u_z+
 {1\over\alpha} u_x+q\|^2_B\cr
&\le 
 \|\gamma_1 u_y+ \gamma_2 u_z\|^2_B+
 2\|\gamma_1 u_y+ \gamma_2 u_z\|_B
 \|{1\over\alpha} u_x+q\|_B+ \|{1\over\alpha} u_x+q\|^2_B\cr
&\le (1+\delta)
 \|\gamma_1 u_y+ \gamma_2 u_z\|^2_B
 +(1+{1\over\delta})\| {1\over\alpha} u_x+q\|^2_B\cr
&\le (1+\delta)\left(\gamma_1^2+\gamma_2^2\right)
(\|u_y\|^2_B+\|u_z\|^2_B)+
(1+{1\over \delta})\|{1\over\alpha}u_x+q\|^2_B.\cr}
$$
Since $\varrho:= \gamma_1^2+\gamma_2^2<1$ we can
choose $\delta$ such that $(1+\delta)\varrho\le (1-\delta).$ Therefore, by (8),
$$
\delta(\|u_y\|^2_B+\|u_z\|^2_B)\le 
(1+{1\over\delta})\| {1\over \alpha} u_x+q\|^2_B +  \|u_x\|^2_B
+2\partt(u_x,u_t) -2 (u_x,F). 
$$

Since $(u_x,u_t)$ can be estimated by $E,$ we have proved
\proclaim
Theorem 2. 
$$
\eqalign{
& \int_0^T \left(\|u_t\|^2_B+\|u_x\|^2_B+\|u_y\|^2_B+\|u_z\|^2_B\right)dt\cr
&\quad \le \con(E(0)+\int_0^T \|F\|^2dt+\int_0^T \|q\|^2_B dt).\cr}
$$
\par
\medskip
The results can easily be generalized to half-plane problems for wave
equations of the general form
$$
u_{tt}=P_0u_t+P_1 u,\quad x_1\ge 0,\quad -\infty < x_j <\infty ,~
j=2,3 \eqno(9)
$$
with boundary conditions
$$
\alpha u_t=u_{x_1}+\beta_1 u_{x_2}+\beta_2 u_{x_3},\quad x_1=0.
\eqno(10) $$
Here
$$
P_0=\sum_{j=1}^3 c_j \ipartx _j ,\quad
P_1=\sum_{i,j=1}^3 a_{ij}\partial^2/\partial x_i\partial x_j,~
a_{ii}>0
$$
and $P_1$ is strongly elliptic.

By general coordinate transformations, we can transform (9),(10) to the simple
problem (1)--(3), except if the coordinate system moves normal to the boundary.
To discuss this case we consider a boundary moving with constant velocity
$c/\sqrt{1+c^2}$
in the $x$-direction and transform (1) according to
$$ x= {x'+ct' \over \sqrt{1+c^2}}, \quad t=t',\quad  y=y', \quad z=z',$$
so that the moving boundary is located at $x'=0$. After dropping the primes,
we obtain
$$ u_{tt}=2cu_{xt}+u_{xx}+u_{yy}+u_{zz}+F \eqno(1') $$ and we consider the problem
(1${}'$),(2),(3).
The same pseudo-differential technique as in Section 2 of [1] shows that this
new problem is well-posed in the generalized sense if and only if
$$
{\alpha +c\over \sqrt{1+c^2}}>\sqrt{\beta_1^2+\beta_2^2}. $$
Subject to this inequality, in Appendix 2 we use a geometric approach to
establish strong well-posedness for this moving boundary problem by the energy
method.

\bigskip
\noindent
{\bf 3. The Einstein equations} 

We consider the half-plane problem treated in [1] for the harmonic Einstein
equations, which we know describe. In the frozen coefficient formalism based
upon an orthonormal frame, the components of the densitized spacetime metric
satisfy the wave equations
$$
\eqalign{ 
&(-{\partial^2 \over \partial t^2}+{\partial^2 \over \partial x^2}
   +{\partial^2 \over \partial y^2} +{\partial^2 \over \partial z^2})
\pmatrix{\gamma^{tt} & \gamma^{tx} & \gamma^{ty} & \gamma^{tz}\cr
\gamma^{tx} & \gamma^{xx} & \gamma^{xy} & \gamma^{xz}\cr
\gamma^{ty} & \gamma^{xy} & \gamma^{yy} & \gamma^{yz}\cr
\gamma^{tz} & \gamma^{xz} & \gamma^{yz} & \gamma^{zz}\cr} =F\cr 
&\quad x\ge 0,~ t\ge 0,~ -\infty < y < \infty,~ -\infty < z < \infty, \cr}
\eqno(11) 
$$
where $F$ consists of lower order terms.

These equations are subject to the constraints
$$ \eqalign{
C^t & =\partt\gamma^{tt}+\partx\gamma^{tx}+\party\gamma^{ty}+\partz\gamma^{tz}=0,\cr
C^x & =\partt\gamma^{tx}+\partx\gamma^{xx}+\party\gamma^{xy}+\partz\gamma^{xz}=0,\cr
C^y & =\partt\gamma^{ty}+\partx\gamma^{xy}+\party\gamma^{yy}+\partz\gamma^{yz}=0,\cr
C^z & =\partt\gamma^{tz}+\partx\gamma^{xz}+\party\gamma^{yz}+\partz\gamma^{zz}=0.
\cr }
\eqno(12)
$$
Because the constraints satisfy homogeneous wave equations, if
$C^a(0,x,y,z)=0$ and ${\partial \over \partial t} C^a(0,x,y,z)=0$, $a=(t,x,y,z)$,
then they remain zero at later times if $ C^a=0$ are part of the boundary
conditions for (11) at $x=0$.

As in [1], we consider the choice of boundary conditions
$$ \eqalign{
\partt 
\pmatrix{
\gamma^{tt}\cr \gamma^{tx}\cr \gamma^{xx}\cr \gamma^{ty}\cr
\gamma^{xy}\cr \gamma^{tz}\cr \gamma^{xz}\cr
\gamma^{yy}\cr \gamma^{yz}\cr \gamma^{zz}\cr} &+\partx
\pmatrix{
0 & 1 & 0 & 0 & 0 & 0 & 0 & 0 & 0 & 0 \cr
0 & 0 & 1 & 0 & 0 & 0 & 0 & 0 & 0 & 0 \cr
a_1 & a_2 & a_3 & 0 & 0 & 0 & 0 & 0 & 0 & 0 \cr
0 & 0 & 0 & 0 & 1 & 0 & 0 & 0 & 0 & 0 \cr
0 & 0 & 0 & b_1 & b_2 & 0 & 0 & 0 & 0 & 0 \cr
0 & 0 & 0 & 0 & 0 & 0 & 1 & 0 & 0 & 0 \cr
0 & 0 & 0 & 0 & 0 & c_1 & c_2 & 0 & 0 & 0 \cr
0 & 0 & 0 & 0 & 0 & 0 & 0 & -1 & 0 & 0 \cr
0 & 0 & 0 & 0 & 0 & 0 & 0 & 0 & -1 & 0 \cr
0 & 0 & 0 & 0 & 0 & 0 & 0 & 0 & 0 & -1 \cr}
\pmatrix{
\gamma^{tt}\cr \gamma^{tx}\cr \gamma^{xx}\cr \gamma^{ty}\cr
\gamma^{xy}\cr \gamma^{tz}\cr \gamma^{xz}\cr 
\gamma^{yy}\cr \gamma^{yz}\cr \gamma^{zz}\cr} \cr
&+\party
\pmatrix{
0 & 0 & 0 & 1 & 0 & 0 & 0 & 0 & 0 & 0 \cr
0 & 0 & 0 & 0 & 1 & 0 & 0 & 0 & 0 & 0 \cr
0 & 0 & 0 & 0 & 0 & 0 & 0 & 0 & 0 & 0 \cr
0 & 0 & 0 & 0 & 0 & 0 & 0 & 1 & 0 & 0 \cr
0 & 0 & 0 & 0 & 0 & 0 & 0 & 0 & 0 & 0 \cr
0 & 0 & 0 & 0 & 0 & 0 & 0 & 0 & 1 & 0 \cr
0 & 0 & 0 & 0 & 0 & 0 & 0 & 0 & 0 & 0 \cr
0 & 0 & 0 & 0 & 0 & 0 & 0 & 0 & 0 & 0 \cr
0 & 0 & 0 & 0 & 0 & 0 & 0 & 0 & 0 & 0 \cr
0 & 0 & 0 & 0 & 0 & 0 & 0 & 0 & 0 & 0 \cr}
\pmatrix{
\gamma^{tt}\cr \gamma^{tx}\cr \gamma^{xx}\cr \gamma^{ty}\cr
\gamma^{xy}\cr \gamma^{tz}\cr \gamma^{xz}\cr 
\gamma^{yy}\cr \gamma^{yz}\cr \gamma^{zz}\cr} \cr
&+\partz
\pmatrix{
0 & 0 & 0 & 0 & 0 & 1 & 0 & 0 & 0 & 0 \cr
0 & 0 & 0 & 0 & 0 & 0 & 1 & 0 & 0 & 0 \cr
0 & 0 & 0 & 0 & 0 & 0 & 0 & 0 & 0 & 0 \cr
0 & 0 & 0 & 0 & 0 & 0 & 0 & 0 & 1 & 0 \cr
0 & 0 & 0 & 0 & 0 & 0 & 0 & 0 & 0 & 0 \cr
0 & 0 & 0 & 0 & 0 & 0 & 0 & 0 & 0 & 1 \cr
0 & 0 & 0 & 0 & 0 & 0 & 0 & 0 & 0 & 0 \cr
0 & 0 & 0 & 0 & 0 & 0 & 0 & 0 & 0 & 0 \cr
0 & 0 & 0 & 0 & 0 & 0 & 0 & 0 & 0 & 0 \cr
0 & 0 & 0 & 0 & 0 & 0 & 0 & 0 & 0 & 0 \cr}
\pmatrix{
\gamma^{tt}\cr \gamma^{tx}\cr \gamma^{xx}\cr \gamma^{ty}\cr
\gamma^{xy}\cr \gamma^{tz}\cr \gamma^{xz}\cr 
\gamma^{yy}\cr \gamma^{yz}\cr \gamma^{zz}\cr} =q,\cr}
\eqno(13)$$
where $q$ represents the boundary data.
Here $a_1,~a_2,~a_3,~b_1,~b_2$ and $c_1,~c_2$ are real constants such that
the eigenvalues $\lambda_j$ of
$$
\pmatrix{ 0 & 1 & 0\cr 0 & 0 & 1 \cr a_1 & a_2 & a_3 \cr},\quad
\pmatrix{0 & 1\cr b_1 & b_2 \cr} \quad {\rm and}\quad
\pmatrix{ 0 & 1 \cr c_1 & c_2 \cr}
\eqno(14) $$
are real and negative. The constraints vanish on the boundary provided
$q^{ta}=0$, $a=(t,x,y,z)$.

We can now obtain standard energy estimates for the Einstein equations both
in the interior and on the boundary. The following theorem strengthens the
results in Section 3 of [1] to strong well-posedness:

\proclaim
Theorem 3. The half-plane problem for the system (11) with constraints (12) and
boundary conditions (13) is strongly well-posed if the eigenvalues of the
matrices (14) are real and negative.
\par
\medskip\noindent
{\it Proof.}
As explained in Appendix 1, it suffices to
show that all the first derivatives are bounded. We start with the last components. Since the boundary conditions do not
contain terms in the tangential direction, Theorems 1 and 2 tell us that all the
first derivatives  of $\gamma^{zz}$ can be bounded in terms of $\|F\|$ and
$\|q\|_B.$ Thus we gain one derivative both on the boundary and in the interior.
The same result holds for the derivatives of $\gamma^{yz}$ and $\gamma^{yy}.$

In the same way as in [1], the boundary conditions for
$\gamma^{tz},\gamma^{xz}$ can be decoupled by a unitary matrix $U$ such that
$$ U \pmatrix{0 & 1 \cr c_1 & c_2 \cr}U^*=
-\pmatrix{ \lambda_1 & c_{12} \cr 0 & \lambda_2 \cr},
\quad \lambda_1,\lambda_2 >0.
$$
Introducing new variables by
$$
\pmatrix{
\tilde\gamma^{tz}\cr \tilde\gamma^{xz}\cr} =
U \pmatrix{
\gamma^{tz}\cr \gamma^{xz}  \cr}
$$
we obtain the equations
$$
{\partial \over \partial t} 
\pmatrix{\tilde\gamma^{tz}\cr \tilde\gamma^{xz}\cr}
=\pmatrix{\lambda_1 &c_{12}\cr 0 & \lambda_2\cr}
{\partial \over \partial x} 
\pmatrix{\tilde\gamma^{tz}\cr \tilde\gamma^{xz}\cr}
+\pmatrix{\tilde q^{tz}\cr \tilde q^{xz}\cr}
$$
where
$$
\pmatrix{\tilde q^{tz}\cr \tilde q^{xz}\cr}=
U\left(
-{\partial \over \partial y} \pmatrix{\gamma^{yz}\cr 0 \cr}
-{\partial \over \partial z} \pmatrix{\gamma^{zz}\cr 0 \cr}
+\pmatrix{ q^{tz}\cr  q^{xz}\cr}\right).
$$
We have already estimates of $\tilde q^{tz}$ and $\tilde q^{xz}.$ Therefore we
can estimate all the first
derivatives of $\tilde\gamma^{xz}$ in terms of $\|F\|$ and $\|q\|_B.$ The same
is true for $\tilde\gamma^{tz}.$ This process can be continued for the
remaining components. Thus we have proved Theorem 3 of [1], where we can remove
``in the generalized sense''.

Theorem 3 is also valid when the matrices of (13) for the tangential
derivatives are upper triangular, i.e. only terms above the diagonal are
not zero (or equivalent to that form by unitary transformation). This
allows the sequential argument in the proof. It also generalizes to full
matrices which are sufficiently close to upper triangular form. A fuller
discussion of the most general case will be given in future work.

Local existence theorems for quasilinear equations follow by iteration of the
linearized equations, as described in Appendix 1. These results establish
strong well-posedness, locally in time, of the quasilinear harmonic IBVP.

\bigskip\noindent
{\bf 4. Discussion}
\bigskip

We have shown that the results of [1], obtained using pseudo-differential
theory, can also be derived in a more transparent way based upon integration by
parts to establish strong well-posedness of a broad class of IBVP's governed by
second order quasilinear wave equations. The underlying arguments require no
need to rewrite the wave equations as a first order system. As shown in Appendix
1, for smooth data there exist estimates for arbitrarily high derivatives (which
is a key requirement for treating the quasilinear case). The boundary conditions
are flexible and are stable against perturbation of their coefficients. These
properties are important for numerical calculations.

We anticipate that these results will have application to the broad class of
problems based upon second order wave equations. For the standard wave equation
(1), the chief restriction is that the coefficient $\alpha$ of $u_t$ in the
boundary condition (2) is positive and satisfies $\alpha^2>\beta_1^2+
\beta_2^2$; for the general wave equation on a curved space background, the
corresponding restriction is that the vector $T^a$ introduced in Appendix 2 is
timelike and future directed. This encompasses boundary conditions of Sommerfeld
type. The application to Maxwell's equations expressed in terms of a vector
potential in the Lorentz gauge is straightforward.

A prime motivation for this work is the formulation of a well-posed IBVP for
the system of harmonic Einstein equations. The importance of numerical
simulations in general relativity has spurred a large number of works which
have established many of the necessary ingredients for a well-posed IBVP. For a
review see [7]. The first complete well-posed formulation was given by
Friedrich and Nagy [8] for a version of Einstein's equations in which the
curvature tensor, i.e. quantities constructed out of second derivatives of the
metric, was included in the evolved variables. Choquet-Bruhat's first proof of
the well-posedness of the Cauchy problem for Einstein's equation was given in
the harmonic formulation [9]. It is satisfying to extend her work to the 
harmonic IBVP. Our results also immediately apply to the generalized harmonic
formulation in which harmonic forcing terms are allowed [10], since this does
not change the principle part of the system. 

Since the pioneering results of Pretorius [11,12], there has been rapid
progress in the development of numerical codes based upon the generalized
harmonic formulation with the capabilty of simulating relativistic binaries
consisting of black holes or relativistic stars [13-15]. This difficult problem
requires extensive computational tools that do not enter the analytic treatment
considered here, namely grid refinement and numerical dissipation.
Nevertheless, our results provide the analytic background necessary to justify
a successful numerical treatment and, at least for finite difference codes,
they provide further guidance on how to establish robust boundary conditions.
Examples of the boundary conditions considered here have been incorporated in a
unigrid, second differential order, harmonic code and successfully tested on
model problems [16]. It is beyond the scope of the present work to suggest how
they might be incorporated in specific black hole codes. However, in future
work we will show how the geometric approach of Appendix 2 can be further
developed to yield boundary conditions which are well tailored to the treatment
of isolated astrophysical systems.


\bigskip\noindent
{\bf Appendix 1. The quasilinear case}
\bigskip

The preceding energy estimates establish that a solution of the IBVP with
frozen coefficients is unique and depends continuously on the data. In this
section we want to show that local existence theorems and energy estimates for
second order quasilinear wave equations are proved in the same way as for first
order symmetric hyperbolic systems. It all depends on a priori estimates for
arbitrarily high derivatives of the solutions of linear equations with variable
coefficients. Consider the halfplane problem for
$$
u_{tt}=Pu+Ru+F,\quad x\ge 0,\quad  -\infty<y<\infty, \eqno(15)
$$
with boundary conditions
$$
\alpha u_t=u_x-\gamma u+ru+q,\quad \alpha(y,z,t) \ge \delta >0, \eqno(16)
$$
and initial data
$$
u(t=0)=f_1,\quad u_t(t=0)=f_2. \eqno(17)
$$
Here
$$ Pu=(au_x)_x+(bu_y)_y-2\gamma u_t-\gamma^2u $$
and
$$ Ru=c_1u_t+c_2u_x+c_3u_y+c_4u.$$
$Ru$ are terms of lower (first and zeroth) differential order. 
All coefficients are smooth functions of 
$x,y,t$ and $a\ge a_0>0,~b\ge b_0>0,~a_0,b_0$ and $\delta$
are strictly positive constants.
The initial data are smooth functions which are compatible with the
boundary conditions.
Here $\gamma>0$ is a constant obtained by the change of variables
$u\to e^{\gamma t}u'$ and then deleting the 'prime'. This
introduces the term $\gamma^2\|u\|^2$ in the energy $E$
in (18), which provides an estimate of $\|u\|^2$.
\proclaim
Lemma 2. There is an energy estimate which is stable against lower
 order perturbations.
\par
\medskip\noindent
{\it Proof.} Integration by parts gives
$$\eqalign{
\partt E&
:=\partt\left( \|u_t\|^2+(u_x,au_x)+(u_y,bu_y)+\gamma^2
\|u\|^2\right)\cr
&= -4\gamma\|u_t\|^2+2(u_t,F)+2(u_t,Ru)-2(u_t,a u_x)_B\cr
     & +a_t \|u_x\|^2+b_t \|u_y\|^2   \cr
&\le \con(\|F\|^2+E)-2(u_t,au_x)_B.\cr}
\eqno(18) $$

\bigskip

Using the boundary conditions gives
$$\eqalign{
-(u_t,au_x)_B&= -(u_t,\alpha au_t)_B-(u_t,\gamma au)_B+(u_t,ru+q)_B\cr
&=-(u_t,\alpha au_t)_B-(u_t,\gamma a_0u)_B-\left(u_t,\gamma(a-a_0)u\right)_B
+(u_t,aru+aq)_B\cr
&\le -\halv \gamma a_0\partt \|u\|^2_B-\halv (u_t,\alpha au_t)_B+
\con (\|u\|^2_B+\|q\|^2_B).\cr}
$$
Therefore, we obtain from (18)
$$ \eqalign{
&\partt (E+\gamma a_0\|u\|^2_B+(u_t,\alpha a u_t)_B )\cr
&\le \con \left(E+\|u\|^2_B+\|F\|^2+\|q\|^2_B\right).\cr}
\eqno(19) $$
This proves the lemma.
\medskip
Now we can estimate the derivatives. Let $v=u_y,w=u_t.$ Differentiating
the differential equation gives us
$$ \eqalign{ 
v_{tt}&=Pv+Rv+R_yu+(a_yu_x)_x+(b_yv)_y+F_y,\cr
w_{tt}&=Pw+Rw+R_tu+(a_tu_x)_x+(b_tv)_y+F_t,\cr}
\eqno(20) $$
respectively.

$R_yu$ and $R_tu$ are linear combinations of first derivatives of $u$
which we have already estimated and can be considered part of the forcing.

The differential equation (15) tells us that 
$$ au_{xx}=w_t-bv_y+~ \hbox{terms we have already estimated.} $$
Thus $u_{xx}$ is lower order with respect to $v,w$ and, except for
lower order terms, $v,w$ are solutions of the same differential equation
as $u.$ The same is true for the boundary conditions. Therefore we can
estimate all second derivatives. Repeating the process, we can estimate
 any number of derivatives.

We can now proceed in the same way as in [2], where we have considered
first order systems, to obtain existence theorems for equations with
variable coefficients. We approximate the differential equation by a stable
difference approximation and prove, using summation by parts, that the
corresponding estimates for the divided differences hold independently of the
gridsize. In the limit of vanishing gridsize, we obtain the existence theorem. 
Since we can estimate any number of derivatives, it is well known, using
Sobolev's theorem, that we can obtain similar, although local in time,
estimates for quasilinear systems. By the same iterative methods as for first
order symmetric hyperbolic systems, it follows that well-posedness extends
locally in time to the quasilinear case, as well as other standard results
such as the principle of {\it finite speed of propagation}.


\bigskip\noindent
{\bf Appendix 2. Geometric Derivation of the Estimates}
\bigskip

We now use a geometric approach to show how the estimates established in Sec. 2
can be extended to the general initial-boundary value problem for the second
order wave equation on a curved spacetime. A similar geometric approach has
been used to implement the second order harmonic formulation of Einstein's
equations as an evolution-boundary code based upon summation by parts [16]. See
also [17] for another application of the geometric approach to the treatment
of boundaries.

Using standard notation of general relativity, we consider the wave equation
$$
   g^{ab}\nabla_a \nabla_b\phi = F
\eqno(21)
$$
for a massless scalar field propagating on a Lorentzian manifold with boundary
of the form $M = [0,T] \times \Sigma$, where $\Sigma$ is a compact,
$3$-dimensional manifold with smooth boundary $\partial\Sigma$, each time-slice
$\Sigma_t = \{ t \} \times \Sigma$ is spacelike and the boundary ${\cal T} =
[0,T] \times \partial\Sigma$ is timelike. Here $\nabla$ denotes the covariant
derivative associated with the spacetime metric $g$ with signature $(-,+,+,+)$.
The initial-boundary value problem consists in finding solutions of (21)
subject to the initial conditions
$$
\left. \phi \right|_{\Sigma_0} = f, \qquad
\left. n^b\nabla_b\phi \right|_{\Sigma_0} = h ,
$$
with Cauchy data $f$ and $h$ on $\Sigma_0$, and the boundary condition
$$
\left[ (T^b + a N^b) \nabla_b\phi \right]_{\cal T} = q
$$
with data $q$ on ${\cal T}$. Here $n^b$ and $N^b$ denote the
future-directed unit vector field to the time-slices $\Sigma_t$ and the outward
unit normal vector field to ${\cal T}$, respectively; $T^b$ is an arbitrary
future-directed timelike vector field which is tangent to the boundary surface
${\cal T}$; and $a > 0$. The motion of the boundary is described geometrically
by the hyperbolic angle $N^b n_b$.

Without loss of generality, we can assume that $T^b$ is normalized such that
$g_{bc} T^b T^c = -1$. A Sommerfeld boundary condition then corresponds to the
choice $a=1$ for which $T^b +N^b$ points in an outgoing characteristic (null)
direction picked out by the geometry. See [16] for a numerical study of the
harmonic Einstein system carried out with such a Sommerfeld condition, with
evolution in the $T^b$ direction. In [16], the choice of matrices governing the
tangential derivatives in the boundary conditions (13) was made for mathematical
simplicity. In future work, we will explore more physically and geometrically
motivated choices.

In order to establish estimates, we introduce the notation $\phi_a =
\nabla_a\phi$ and the energy momentum tensor of the scalar field  
$$
        \Theta_b^a = \phi_b \phi^a -{1\over 2}\delta_b^a \phi^c \phi_c.
$$
The essential idea is the use of an energy associated with a timelike vector
$u^a=T^a+\delta N^a$, where $\delta>0$,  so that $u^a$
points outward from the boundary.  The corresponding energy $E(t)$ and the
energy flux ${\cal F}(t)$ through the boundary $\Sigma_t$ are
$$
     E(t) =\int\limits_{\Sigma_t} u^b \Theta_b^a n_a 
$$
and 
$$
    {\cal F}(t) =\int\limits_{\partial \Sigma_t} u^b \Theta_b^a N_a .
$$
It follows from the timelike property of $u^a$  that $E(t)$ is a norm for
$\phi_a(t)$.

Energy conservation for the scalar field, i.e. integration by parts, gives
$$
     \partial_t E =  {\cal F}
          -\int\limits_{\Sigma} (\Theta_{ab}\nabla^a u^b +u^a \phi_a F) 
$$
so that
$$
     \partial_t E \le  {\cal F} + \con
         (E+ \int\limits_{\Sigma} F^2 ) .
\eqno(22)
$$

The required estimates arise from considering the flux density
$$
\eqalign{
&   u^b \Theta_b^a N_a =  N^a \phi_a T^b \phi_b +\delta (N^a \phi_a )^2
               -{\delta\over 2}\phi^a \phi_a  \cr 
&=  N^a \phi_a T^b \phi_b +{\delta\over 2} (N^a \phi_a )^2
            + {\delta\over 2} (T^a \phi_a )^2  
             -{\delta\over 2}H^{ab}\phi_a \phi_b  \cr
&= -{\delta\over 2} \left (
       (N^a \phi_a )^2 + (T^a \phi_a )^2  + H^{ab}\phi_a \phi_b \right)
          +N^a \phi_a T^b \phi_b +\delta (N^a \phi_a )^2
            + \delta(T^a \phi_a )^2  
             \cr }
$$
where $H_{bc} = g_{bc} + T_b T_c - N_b N_c$ is the positive definite metric
in the tangent space of the boundary orthogonal to $T^a$. By using
the boundary condition to eliminate $T^a \phi_a$ in the last group of terms, we
obtain
$$
\eqalign{
  u^b \Theta_b^a N_a =  -{\delta\over 2} \left (
       (N^a \phi_a )^2 + (T^a \phi_a )^2  + H^{ab}\phi_a \phi_b \right) \cr
       +\left (-a+\delta(1+a^2)\right)(N^a \phi_a )^2 
      +(1-2a\delta)N^a \phi_a q +\delta q^2 
 \cr }
$$
so that
$$
\eqalign{
   u^b \Theta_b^a N_a =  -{\delta\over 2} \left (
  (N^a \phi_a )^2 + (T^a \phi_a )^2  + H^{ab}\phi_a \phi_b \right) \cr
 +\left (-a\left(1-\epsilon \right)+\delta\left (1+a^2\right ) \right )\left(N^a \phi_a \right)^2 \cr
  -\epsilon a \left (N^a \phi_a -{\left(1-2a\delta \right)\over 2a\epsilon} q\right )^2  
+\left (\delta+ {(1-2a \delta)^2 \over 4a\epsilon}\right)q^2
                  .
\cr }
$$
In the above equation, we have introduced the $\epsilon$-terms,
with $0\le\epsilon\le 1$, in order to
establish the inequality  
$$\eqalign{
  u^b \Theta_b^a  N_a \le  -{\delta\over 2} \left (
       (N^a \phi_a )^2 + (T^a \phi_a )^2  + H^{ab}\phi_a \phi_b \right) \cr
       + \left (-a(1-\epsilon)+\delta(1+a^2) \right )(N^a \phi_a)^2    \cr
     +\left (\delta+ {(1-2a \delta)^2\over 4a\epsilon}\right )q^2
                                                . \cr }
\eqno(23)
$$

The boundary estimate of $\phi_a$ now follows
by requiring $-a(1-\epsilon) +\delta (1+a^2) \le 0$, which guarantees that
$u^a$ is timelike.
With the choice
$$
   \delta ={a(1-\epsilon)\over (1+a^2)},
$$
(22) and (23) give
$$\eqalign{
  \partial_t E + \int\limits_{\partial\Sigma}{\delta\over 2} \bigg(
       (N^a \phi_a )^2 + (T^a \phi_a )^2  + H^{ab}\phi_a \phi_b \bigg) \cr
    \le  \con \bigg (E+ \int\limits_{\Sigma} F^2 +  
       \int\limits_{\partial\Sigma} q^2     \bigg ). 
\cr }
\eqno(24)
$$
In analogy with (19), this leads to the required estimate of the gradient
$\phi_a$ on the boundary (as well as the usual estimate of the gradient at a
fixed time) to prove that the problem is strongly well-posed. An estimate of
$\phi$ itself follows by introducing a mass term in (21) through the change of
variable $\phi\to e^{\gamma t}\phi'$, as described in Appendix 1. Energy
estimates for the problem (1${}'$),(2),(3) in Sec. 2 follow from the choice
$T^b=(T^t,T^x,T^y,T^z)=\tau(\alpha',0,-\beta'_1,-\beta'_2)$ with $N^b=-\nu
g^{ba}\nabla_a x =-\nu(c,1,0,0)$, where $\tau$ and $\nu$ are positive
normalization constants and $\alpha'$ and $\beta'_i$ are related to the
coefficients $\alpha$ and $\beta_i$ in (2) by
$$\eqalign{
   \alpha={\tau \alpha'\over  a\nu}-c \cr
   \beta_i={\tau\beta'_i\over a\nu}, \quad i=1,2. \cr }
$$
\bigskip\noindent

\vfil\eject

\centerline {Acknowledgments}
\bigskip

The work of O. R. was supported in part by CONICET, SECYT-UNC and NSF Grant
INT0204937 to Louisiana State University. The work of O. S. was supported in
part by grant CIC 4.20 to Universidad Michoacana. The work of J. W. was
supported by NSF grant PH-0553597 to the University of Pittsburgh.

\bigskip

\centerline{References}

\bigskip\noindent
[1] H.-O. Kreiss and J. Winicour, {\it Class. Quantum Grav.} {\bf 23},
405--420, (2006).
\smallskip\noindent
[2] H.-O. Kreiss and J. Lorenz, {\bf Initial-Boundary Value Problems and the
Navier-Stokes Equations}, 1989, Reprint {\it SIAM CLASSICS}, 2004.
\smallskip\noindent
[3] B. Gustafsson, H.-O. Kreiss and J. Oliger,
{\bf Time dependent problems and difference methods},
(A Wiley-Interscience Publication, 1995).
\smallskip\noindent
\smallskip\noindent
\smallskip\noindent
[4] H.-O. Kreiss, N.~A. Petersson,
{\it SIAM J. Sci. Comput.} {\bf 27}, 1141 (2006).
\smallskip\noindent
[5] H.-O. Kreiss, N.~A. Petersson, and J. Ystr\"{o}m,
{\it SIAM J. Numer. Anal.} {\bf 42}, 1292 (2004).
\smallskip\noindent
[6] M.~C. Babiuc, B. Szil\'{a}gyi and J. Winicour,
{\it Phys.Rev.} {\bf D73}, 064017 (2006).
\smallskip\noindent
[7] O. Sarbach, {\bf Absorbing boundary conditions for Einstein's
field equations}, arXiv:0708.4266 [gr-qc]
\smallskip\noindent
[8] H. Friedrich and G. Nagy, {\it Commun. Math. Phys.} {\bf 201}, 619 (1999).
\smallskip\noindent
[9] Y. Foures-Bruhat, {\it Acta Math.} {\bf 88}, 141 (1952).
\smallskip\noindent
[10] H. Friedrich, {\it Class. Quantum Grav.} {\bf 13}, 1451 (1996).
\smallskip\noindent
[11] F. Pretorius, {\it Phys. Rev. Lett.} {\bf 95}, 121101 (2005).
\smallskip\noindent
[12] F. Pretorius, {\it Class. Quantum Grav.} {\bf 23}, S529 (2006).
\smallskip\noindent
[13] L. Lindblom, M. Scheel, L.~E. Kidder, R. Owen and O. Rinne,
{\it Class. Quantum Grav.} {\bf 23}, S447 (2006).
\smallskip\noindent
[14] C. Palenzuela, I. Olabarrieta, L. Lehner and S. Liebling,
{\it Phys.Rev.} {\bf D75}, 064005 (2007).
\smallskip\noindent
[15] B. Szil{' a}gyi, D. Pollney, L. Rezzolla, J. Thornburg
and J. Winicour, {\it Class. Quantum Grav.} {\bf 24}, S275 (2007).
\smallskip\noindent
[16] M.~C. Babiuc, H.-O. Kreiss and J. Winicour, {\it Phys. Rev.}
{\bf D75}, 044002 (2007).
\smallskip\noindent
[17] G. Calabrese, L. Lehner, D. Neilsen, J. Pullin, O. Reula, O. Sarbach,
and M. Tiglio, {\it Class. Quantum Grav.} {\bf 20},  L245 (2003).

\bye